\documentclass[onecolumn,showpacs,preprintnumbers,amsmath,amssymb,11pt]{revtex4}

\usepackage{amssymb}
\usepackage{hyperref}

\begin{document}
\title{Quasinormal modes in pure de sitter spacetimes}
\author{Da-Ping Du}
\email{022019004@fudan.edu.cn}
 \author{Bin Wang}
\email{binwang@fudan.ac.cn}
\address{Department of Physics, Fudan University, Shanghai 200433, P.R.China}
\author{Ru-Keng Su}
\email{rksu@fudan.ac.cn}
\address{China Center of Advanced Science and Technology, World Lab, P. O. Box 8730, 100080 Beijing, and Department of Physics, Fudan University, Shanghai 200433, P. R. China}  

\begin{abstract}
We have studied scalar perturbations as well as fermion
perturbations in pure de Sitter space-times. For scalar
perturbations we have shown that well-defined quasinormal modes in $d$-dimensions
can exist provided that the mass of scalar field
$m>\frac{d-1}{2l}$. The quasinormal modes of fermion
perturbations in three and four dimensional cases have also been
investigated. We found that different from other dimensional cases,
in the three dimensional pure de Sitter spacetime there is no
quasinormal mode for the s-wave. This interesting difference
caused by the spacial dimensions is true for both scalar and
fermion perturbations.
\end{abstract}

\maketitle

\section{Introduction}

It is well known that the surrounding geometry of a black hole
will experience damped oscillations under perturbations. The
frequencies and damping times of the oscillations are entirely
fixed by the black hole parameters and independent of the initial
perturbations. These oscillations are called ``quasinormal
modes"(QNM), which is believed as a characteristic ``sound" of
black holes and would lead to the direct identification of the
black hole existence through gravitational wave observation to be
realized in the near future\cite{1}. Due to the potential
astrophysical interest, a great deal of effort has been devoted to
the study of black holes' QNMs. Most of these studies were
concerned with black holes immersed in an asymptotically flat
spacetime\cite{2}. Considering the case when the black hole is immersed in an expanding universe, QNMs of black holes in de Sitter(dS)
space have also attracted much attention\cite{3}.

Motivated by the discovery of
the AdS/CFT correspondence, the investigation of QNM in anti-de
Sitter(AdS) spacetimes became appealing in the past several years.
It was argued that the QNMs of AdS black holes have direct
interpretation in term of the dual conformal field
theory(CFT)\cite{4}-\cite{9}. In dS space the relation between
bulk dS spacetime and the corresponding CFT at the past boundary
$\mathcal{I}^-$ and future boundary $\mathcal{I}^+$ in the
framework of scalar perturbation spectrums has also been
discussed\cite{10}. A quantitative support of the dS/CFT
correspondence was provided.

Recently QNMs in asymptotically
flat spaces have acquired further attention, since the possible
connection between the classical vibrations of a black hole
spacetime and various quantum aspects was proposed by relating the
real part of the QNM frequencies to the Barbero-Immirzi(BI)
parameter, a factor introduced by hand in order that loop quantum
gravity reproduces correctly the black hole entropy\cite{11}. In
order to see whether this quantum connection is true in
Schwarzschild dS (SdS) spacetime, a number of extentions have been
made\cite{12}\cite{13}\cite{14}\cite{15}. For the nearly
extreme SdS black hole, like the Schwarzschild spacetime, the real
part is found still proportional to the black hole surface
gravity, but instead of the integer $n$ labelling the modes, with
proportional coefficients $\sqrt{\ell(\ell+1)-\frac{1}{4}}$ for
scalar and electromagnetic perturbations and
$\sqrt{(\ell+2)(\ell-1)-\frac{1}{4}}$ for gravitational
perturbations. It is too early to confirm the recent conjecture by
relating the $\ell$-dependent real part to the BI
parameter\cite{12}. The imaginary part of the QNM frequencies for
the nearly extreme SdS black hole was found having an equally
spacial structure with the level spacing depending on the surface
gravity of the black hole, which agrees to the result in
asymptotically flat spacetime and is independent of whether the
perturbation is scalar, electromagnetic or gravitational\cite{12}.
This result was confirmed even for the very small SdS black
holes\cite{13} and later further supported by using the Born
approximation\cite{14}.

The dependence of the surface gravity for both the real part and
imaginary part of QNM frequencies suggests that there is a
possible connection between the QNMs and thermodynamics of the black hole
horizon. But what is the effect of the cosmological horizon
here? It is believed that cosmological horizon has very similar
thermodynamical behavior to that of the black hole horizon \cite{16}. Is there connection between the
QNMs and surface gravity of the cosmological horizon? In order to
answer this question, in this paper we are going to investigate
the pure dS spacetime. It was argued that for a massless minimally
coupled scalar field, there exists no QNMs in the pure dS
spacetimes, however for a massive scalar field, there do exist
well-defined QNMs\cite{14}. We will examine this arguement in
different topological pure dS spacetimes by considering scalar
perturbation and fermion perturbation. Different from the purpose
to test the dS/CFT correspondence\cite{10}, we here concentrate
on the region within the cosmological horizon.

The structure of the paper is as follows: In Sec.II, we will study
the scalar perturbation in d-dimensional pure dS spacetime with a
spherically symmetric space. The extension to topological pure dS
spacetimes will be shown in Sec.III.  In Sec.IV, we will
present the discussion of the fermion perturbation. Our main
results will be summarized in Sec.V.

\section{Scalar perturbation in pure de Sitter space}
The static metric of a $d$ dimensional dS space reads
\begin{equation}\label{2.1}
ds^2=-f(r)dt^2+f^{-1}(r)dr^2+r^2d\Omega_{d-2}^2\;,
\end{equation}
where $f(r)=1-r^2/l^2$($l$ is the minimal radius of dS space) and
$r^2 d\Omega_{d-2}^2$ represents the metric on the $d-2$
dimensional sphere $\textrm{S}^{d-2}$ of radius $r$.

We begin our discussion with a massive scalar field $\Phi$,
 satisfying the Klein-Gordon equation
\begin{equation}\label{2.2}
   {\Phi^{;\nu}}_{;\nu}=m^2\Phi
\end{equation}
 This equation can be separated by
$\Phi=(u(r)/r^{\frac{d-2}{2}})e^{-i\omega t}Y_\ell(\Omega_{d-2})$.
Here the spherical harmonic $Y_\ell(\Omega_{d-2})$ is the eigen
function of $d-2$ dimensional Laplace-Beltrami operator
$\nabla^2_{d-2}$ with the eigenvalue $-\ell(\ell+d-3)$. Using the
tortoise coordinate, $r_*=\int
\textrm{d}r/f(r)=l\tanh^{-1}(r/l)$, we can write the radial part
into a Schr\"{o}dinger-like equation
\begin{equation}\label{2.3}
   -\frac{d^2u}{{dr_*}^2}+V(r_*)u=\omega^2u
\end{equation}
with effective potential
\begin{equation}\label{2.4}
   V(r_*)=-\frac{\mathcal{A}}{l^2\cosh^2(r_*/l)}+\frac{\mathcal{B}}{l^2\sinh^2(r_*/l)}
\end{equation}
 where $\mathcal{A}$ and $\mathcal{B}$ are defined by
\begin{eqnarray*}
 \mathcal{A} &=& \frac{d-2}{2}(\frac{d-2}{2}+1)-m^2l^2, \\
  \mathcal{B}&=& \ell(\ell+d-3)+\frac{d-2}{2}(\frac{d-2}{2}-1).
\end{eqnarray*}

When $\mathcal{B}>0$, the effective potential (\ref{2.4}) diverges to a positive infinity
at the origin($r=0$) and vanishes at the cosmological
horizon. On the other hand, if $\mathcal{B}<0$, the
potential falls down to negative infinity at the origin which indicates the instability
of the perturbation. We will show it in detail later.

In the form of a new variable $z=1/\cosh^2(r_*/l)$, equation
(\ref{2.3}) becomes
\begin{equation}\label{2.5}
    z(1-z)u''+(1-\frac{3}{2}z)u'+\frac{1}{4}\Big[\frac{\omega^2l^2}{z}-\frac{\mathcal{B}}{1-z}+\mathcal{A}\Big]u=0.
\end{equation}
Further using the ansatz $u=z^\alpha(1-z)^\beta F(z)$, we have
\begin{eqnarray}
  &&
  z(1-z)F''+\Big[1+2\alpha-(2\alpha+2\beta+\frac{3}{2})z\Big]F'+\Big\{\frac{1}{z}\Big(\alpha^2+\frac{\omega^2l^2}{4}\Big)\nonumber\\
  &&\qquad+\frac{1}{1-z}\Big(\beta^2-\frac{1}{2}\beta-\frac{\mathcal{B}}{4}\Big)-\Big((\alpha+\beta)^2+\frac{1}{2}(\alpha+\beta)-\frac{\mathcal{A}}{4}\Big)\Big\}F=0  \label{2.6}
\end{eqnarray}
If we properly select values of $\alpha,\beta$ to make terms like
$\frac{1}{z}$ and $\frac{1}{1-z}$ disappear, the solution to
equation(\ref{2.6}) is the standard hypergeometric function. It is
not hard to find that $z^{\alpha}\sim\exp(\pm\; i\omega r_*)$ when
$z\rightarrow 0$(that is, approaching the horizon), so that the
two independent solutions exactly correspond to the incoming and
outgoing waves at the cosmological horizon.  The general solution is
\begin{eqnarray}
  &&u(z) = u_{1}(z)+u_{2}(z) \nonumber\\
  &&\qquad =C_{1}z^{-\alpha}(1-z)^\beta\
  _2F_1(a-c+1,b-c+1,2-c,z)\nonumber\\
  &&\qquad +\,C_{2}z^{\alpha}(1-z)^\beta\ _2F_1(a,b,c,z)
\end{eqnarray} \label{2.7}
where $C_{1},C_{2}$ are constants, and the parameters of the
hypergeometric function are given by
\begin{eqnarray*}
  c &=& 2\alpha+1 \nonumber\\
  a &=& \alpha+\beta+\frac{1}{4}(1+\sqrt{1+4\mathcal{A}}) \\
  b &=& \alpha+\beta+\frac{1}{4}(1-\sqrt{1+4\mathcal{A}})
  \nonumber
\end{eqnarray*}
Note that the condition of writing the solution in the form of (7) is that $c$ is not an integer.

We will restrict ourselves in the well-accepted definition of the
quasinormal mode which is, in this special case, that these modes
are purely outgoing waves at
 the cosmological horizon and vanish at $r=0$. Indeed this boundary condition is determined by the behavior of the effective potential. Since the sign of $\alpha$ is arbitrary, it is easy to check that  choosing either $u_1$ or $u_2$ as the outgoing wave does't lead to any difference. Here we take $C_1=0$(so $\alpha=-i\omega l/2$), hence the incoming wave is eliminated.
According to the property of hypergeometric function\cite{26}, we
can change the wave function into
\begin{eqnarray}
  &&u_2(z) =C_2\ z^\alpha(1-z)^\beta
   \frac{\Gamma(c)\Gamma(c-a-b)}{\Gamma(c-a)\Gamma(c-b)}\ _2F_1(a,b,a+b-c+1,1-z)\nonumber\\
  &&+C_2\ z^\alpha(1-z)^{\frac{1}{2}-\beta}\frac{\Gamma(c)\Gamma(a+b-c)}{\Gamma(a)\Gamma(b)}\
  _2F_1(c-a,c-b,c-a-b+1,1-z) \label{2.8}
\end{eqnarray}
To make it zero at the origin, we use the poles of Gamma functions.
And to obtain the discrete poles which result in the level spacing
frequencies we should assume that
\begin{equation*}
  \beta(\frac{1}{2}-\beta)=-\frac{\mathcal{B}}{4}\leq 0,
\end{equation*}
which is identical to our previous analysis of the effective
potential. It is interesting to find that in three dimensional case, this condition is equivalent to $l^2\geq 1/4$, which implies that there exists no QNM for s-wave perturbation.

There are two sets of poles. (i) When
 $\beta=\frac{d+2\ell}{4}-\frac{1}{2}\geq\frac{1}{2}$  (note that for three dimensional case we
have excluded the s-wave perturbation for the reason mentioned), the poles are $a=-n$ or $b=-n$
($n=0,1,\cdots$). These poles indeed remove the divergent part of
the wave function around the origin since both $c$ and
$a+b-c=\frac{d+2\ell-3}{2}$ could not be nonpositive integers and
the numerator remains regular.
(ii)When $\beta=-\frac{d+2\ell}{4}+1<0$, the poles are $c-a=-n$ or
$c-b=-n$ ($n=0,1,\cdots$). Because both $c$ and
$c-a-b=\frac{d+2\ell-3}{2}$ could not be nonpositive integers,
these poles are well defined.

The corresponding frequencies are
\begin{eqnarray}
  \omega &=& -\frac{i}{l}(2n+\ell+h_\pm) \label{2.9}\\
  \textrm{or}\qquad\, \omega&=& -\frac{i}{l}(2n-\ell-d+3+h_\pm)\label{2.10}
\end{eqnarray}
 where
$h_\pm=\frac{d-1}{2}-\sqrt{(\frac{d-1}{2})^2-m^2l^2}$.

The above discussion should be restricted to the condition that $c$ is not an integer.
Now let us see what will happen when this
condition is violated. It only occurs in the massless case $m=0$.
We assume that $c=k$, where $k=0,\pm 1, \pm 2, \cdots$. When
$c=k\geq 1$, we can obtain the solution satisfying the boundary
condition\cite{26}
\begin{eqnarray}
   u(z)&\propto&\; z^{-\frac{i\omega l}{2}}(1-z)^{\frac{d+2\ell-2}{4}}\ _2F_1(a,b,k,z)\label{2.11}
\end{eqnarray}
and the corresponding quasinormal frequencies are
\begin{equation}\label{2.12}
    \omega=\frac{i}{l}(k-1)
\end{equation}
When $c=k\leq 0$, the proper solution is then\cite{26}
\begin{eqnarray}
 u(z) &\propto& z^{-\frac{i\omega l}{2}}(1-z)^{\frac{d+2\ell-2}{4}}\
  _2F_1(a-k+1,b-k+1,2-k,z)\label{2.13}
\end{eqnarray}
and the quasinormal frequencies are
\begin{equation}\label{2.14}
\omega=-\frac{i}{l}(k-1).
\end{equation}

 It is surprising that all these
modes (\ref{2.9}) (\ref{2.10})(\ref{2.12})(\ref{2.14}) are just
some kinds of ``distribution" along the radius. To show that, we
compute its flux\cite{7}
\begin{equation}\label{2.15}
    \mathcal{F}=\sqrt{|g|}\frac{1}{2i}(R^*\partial_r R-R\partial_r R^*)
\end{equation}
where $R=u_2(r)/r^{\frac{d-2}{2}}$ is the radial factor of the
wave function. If frequencies $\omega$ in (\ref{2.9})
(\ref{2.10})(\ref{2.12})(\ref{2.14}) are all purely imaginary
values, $\alpha=-i\omega l/2$ is real and so are the
parameters $a$, $b$ and $c$. Then $R(r)$ is
proportional to a purely real function defined in the region $0\leq r\leq l$,
which results in the vanishing of the flux (\ref{2.15})
everywhere as well as at the horizon! This is in contradiction
with the definition of QNM.

However for the massive case, one can see from the scalar
perturbation spectrum (\ref{2.9})(\ref{2.10})
 that if $m^2l^2>(\frac{d-1}{2})^2$, the frequencies are not purely imaginary values which insure that the flux does not vanish. These modes are purely
 outgoing waves
at the cosmological horizon and so are well-defined QNMs.
Therefore there exists the lowest bound ($m>(d-1)/2l$) of the mass
of scalar field that permits QNMs to survive. We rewrite the
corresponding QNM frequencies as follows:
\begin{eqnarray}
  \omega &=& \pm\frac{1}{l}\big[m^2l^2-(\frac{d-1}{2})^2\big]^{\frac{1}{2}}-\frac{i}{l}(2n+\ell+\frac{d-1}{2}) \label{2.16}\\
  \textrm{or}\qquad\,\omega &=& \pm\frac{1}{l}\big[m^2l^2-(\frac{d-1}{2})^2\big]^{\frac{1}{2}}-\frac{i}{l}(2n-\ell+3-\frac{d+1}{2})
  \label{2.17}
\end{eqnarray}

This result confirms the argument that QNMs cannot exist for
massless scalar field\cite{14}. Here we further present the lowest
mass bound for the scalar field to posses QNMs.

\section{scalar perturbation in topological dS space}

In AdS spacetime, the QNMs of different topological black
holes have been studied recently. It was found that black hole
topology influence a lot on the QNMs of scalar
perturbations\cite{8}. Here we would like to extend the
investigation to the topological dS space.

The metric of the topological dS space is showed as (\ref{2.1})
with\cite{18}
\begin{equation}\label{3.1}
    f(r)=\textit{k}+\frac{2Gm}{r^{d-3}}-\frac{r^2}{l^2}
\end{equation}
where $\textit{k}=0,-1$ correspond to two kinds of hypersurfaces
$\Omega_{d-2}$: Ricci flat space $R^{d-2}$ and negative constant
curvature space $H^{d-2}$, respectively. Note that $d\geq 4$ since $d=3$ is
trivial.

The wave equation for the scalar perturbation still has the form
as (\ref{2.3}), but the effective potential
\begin{equation}\label{3.2}
V(r)=\frac{K_{\Omega}}{r^2}+\frac{d-2}{2}\frac{ff'}{r}+\frac{d-2}{2}(\frac{d-2}{2}-1)\frac{f^2}{r^2}+m^2f(r)
\end{equation}
where $K_\Omega$ is the eigenvalue of Laplace-Beltrami operator
$\nabla^2_{d-2}$. It is easy to find that the effective potential
tends to zero at cosmological horizon, while it goes to negative
infinity when $r \rightarrow 0$. Due to the existence of negative
effective potential for the scalar field equation of motion, the
bound states can be formed leading to the growing modes instead of
decaying modes\cite{19}.

The growing modes imply that these topological dS space are not
stable. This result is not so surprising, since both of these
topological dS spaces possess naked singularity.

In AdS case, though the wave amplification behavior exists in
hyperbolic space, for flat space it still exhibits decaying wave
for scalar perturbation\cite{8}. The growing modes for the flat
hypersurface in dS case can be considered as one more difference
between dS and AdS spacetimes.

\section{Fermion perturbation}

\subsection{Three dimensional case}
The fermion perturbation in a BTZ background was studied in
\cite{22,25,7}. We are interested to generalize the investigation
of the fermion perturbation in a pure dS space. There is a little
difference between three and four dimensional case due to the
different representations of gamma matrixes. We first consider the
three dimensional case. After the coordinate transformation
$t=\tau$, $r=l\cos\mu$ and $\phi=\varphi/l$, metric (\ref{2.1})
becomes
\begin{equation}\label{4A.1}
    \textrm{d}s^2=-\sin^2\mu \,\textrm{d}\tau^2+l^2\textrm{d}\mu^2+\cos^2\mu
    \,\textrm{d}\varphi^2.
\end{equation}

We begin with the Dirac equation
\begin{equation}\label{4A.2}
    \gamma^a {e_a}^\nu(\partial_\nu+\Gamma_\nu)\Psi+m\Psi=0,
\end{equation}
where $\gamma^a$ are the conventional spin matrixes. In three
dimensional case we choose the representation of them in term of
Pauli matrixes: $\gamma^0=i\sigma^2$, $\gamma^1=\sigma^1$,
$\gamma^2=\sigma^3$. The triads ${e^a}_\nu$ are
\begin{eqnarray}\label{4A.3}
{e^0}_\tau=\sin\mu,&{e^1}_\mu=l,&{e^2}_\varphi=\cos\mu.
\end{eqnarray}
The spin connection, defined by
$\Gamma_\nu=\frac{1}{8}\gamma^{[a}\gamma^{b]}{e_a}^\lambda
e_{b\lambda;\nu}$, has only two non-vanishing components
\begin{eqnarray}\label{4A.4}
  \Gamma_\tau=-\frac{1}{4l}\cos\mu\,\gamma^{[0}\gamma^{1]}, &\quad&
  \Gamma_\varphi=-\frac{1}{4l}\sin\mu\,\gamma^{[2}\gamma^{1]}.
\end{eqnarray}
We write out equation(\ref{4A.2}) explicitly
\begin{equation}\label{4A.5}
   \Big[\frac{1}{l}\sigma^1(\partial_\mu+\frac{\cos\mu}{2\sin\mu}-\frac{\sin\mu}{2\cos\mu})+\frac{i\sigma^2}{\sin\mu}\partial_\tau+\frac{\sigma^3}{\cos\mu}\partial_\varphi+m\Big]\Psi=0.
\end{equation}
Separating the equation by
\begin{eqnarray*}\label{4A.6}
   \Psi&=&\frac{e^{-i\omega\tau}e^{-i\ell\;\varphi/l}}{\sqrt{\sin\mu\;\cos\mu}}\begin{pmatrix}
     \Psi_1 \\
     \Psi_2 \\
   \end{pmatrix},
\end{eqnarray*}
we arrive at the radial part equation
\begin{eqnarray}
  (\partial_\mu+\frac{i\omega l}{\sin\mu})\Psi_1 &=& -(m l+\frac{i\ell}{\cos\mu})\Psi_2, \label{4A.7}\\
  (\partial_\mu-\frac{i\omega l}{\sin\mu})\Psi_2 &=& -(m
  l-\frac{i\ell}{\cos\mu})\Psi_1, \label{4A.8}
\end{eqnarray}
Introducing a new set of wave functions $\psi_{1},\psi_2$, which
relate to $\Psi_1,\Psi_2$ by
\begin{eqnarray}
  \Psi_1+\Psi_2 &=& (1+\tan^2\mu)^{-\frac{1}{4}}\sqrt{1+i\tan\mu}\;(\psi_1+\psi_2) \label{4A.9}\\
  \Psi_1-\Psi_2 &=&
  (1+\tan^2\mu)^{-\frac{1}{4}}\sqrt{1-i\tan\mu}\;(\psi_1-\psi_2)
  \label{4A.10}
\end{eqnarray}
and using a new variable $y=\tan\mu$ for the sake of convenience,
we turn Dirac equations into
\begin{eqnarray}
  &&(1+y^2)\partial_y\psi_1+(-\ell y+\frac{il\omega}{y})\psi_1 = -\Big[(m l+\frac{i}{2})+(i\ell-l\omega)\Big]\psi_2 \label{4A.11},\\
 &&(1+y^2)\partial_y\psi_2-(-\ell y+\frac{il\omega}{y})\psi_2 = -\Big[(m
 l+\frac{i}{2})-(i\ell-l\omega)\Big]\psi_1\label{4A.12}.
\end{eqnarray}
Further by setting $z=-y^2$ and choosing the ansatz
$\psi_1=z^\alpha(1-z)^\beta F(z)$, from the coupled equation
(\ref{4A.11})(\ref{4A.12}), we find the purely outgoing solutions
in terms of hypergeometric function:
\begin{eqnarray}
  &&  \psi_1=B_1 z^{\alpha}(1-z)^\beta\ _2F_1(a,b,c,z), \label{4A.13}\\
  &&
  \psi_2=B_2z^{\frac{1}{2}+\alpha}(1-z)^{\beta}\Big[(\frac{\ell}{2}-\alpha+\beta)\;_2F_1(a,b,c,z)\nonumber\\
  &&\qquad\qquad+\frac{ab}{c}(1-z)\,\,_2F_1(a+1,b+1,c+1,z)\Big],\label{4A.14}
\end{eqnarray}
where the constants $\alpha,\beta$ ,the hypergeometric parameters
$a,b,c$ and coefficients $B_1,B_2$ are
\begin{eqnarray}
  \alpha &=& -\frac{i\omega l}{2},\nonumber\\
  \beta &=& \pm (\frac{1}{4}-\frac{i m l}{2}),\nonumber\\
  a &=& -\frac{\ell}{2}+\alpha+\beta ,\nonumber\\
  b &=& \frac{1+\ell}{2}+\alpha+\beta ,\label{4A.15}\\
  c &=& \frac{1}{2}+2\alpha ,\nonumber\\
  B_1&=&B_2(\frac{\ell}{2}-\alpha\pm\beta).\nonumber
\end{eqnarray}

We now consider the flux along the radial direction
\begin{eqnarray}\label{4A.16}
   &&\mathcal{F}=\sqrt{|g|}{e_a}^\mu\Psi^+\gamma^a\Psi=(|\psi_1|^2-|\psi_2|^2).
\end{eqnarray}
Using the property of hypergeometric function\cite{26}, we can
obtain the asympotic behavior of $\psi_1$ when $z\rightarrow
-\infty$
\begin{eqnarray}
  &&\psi_{1} =B_1(1-\frac{1}{z})^{\beta}(-1)^\alpha \Big\{ \nonumber\\
  &&\, (-z)^{\alpha+\beta-a}\;\frac{\Gamma(c)\Gamma(b-a)}{\Gamma(c-a)\Gamma(b)}\ _2F_1(a,a-c+1,a-b+1,\frac{1}{z}) \nonumber\\
   && +\ (-z)^{\alpha+\beta-b}\;
   \frac{\Gamma(c)\Gamma(a-b)}{\Gamma(c-b)\Gamma(a)}\
   _2F_1(b,b-c+1,b-a+1,\frac{1}{z})\Big\}\nonumber\\
   &&\simeq B_1(-1)^\alpha \Big[(-z)^{\alpha+\beta-a}\;\frac{\Gamma(c)\Gamma(b-a)}{\Gamma(c-a)\Gamma(b)}+(-z)^{\alpha+\beta-b}\;
   \frac{\Gamma(c)\Gamma(a-b)}{\Gamma(c-b)\Gamma(a)}\Big].\label{4A.17}
\end{eqnarray}
Similarly for $\psi_2$ we have
\begin{eqnarray}
  &&\psi_{2} \simeq B_2(-1)^{\alpha+\frac{1}{2}}\Big\{ \nonumber\\
  &&(\frac{\ell}{2}-\alpha+\beta)\Big[(-z)^{\frac{1}{2}+\alpha+\beta-a}\;\frac{\Gamma(c)\Gamma(b-a)}{\Gamma(c-a)\Gamma(b)}+(-z)^{\frac{1}{2}+\alpha+\beta-b}\;
   \frac{\Gamma(c)\Gamma(a-b)}{\Gamma(c-b)\Gamma(a)}\Big] \nonumber\\
   &&  +\frac{ab}{c}\Big[(-z)^{\frac{1}{2}+\alpha+\beta-a}\;\frac{\Gamma(c+1)\Gamma(b-a)}{\Gamma(c-a)\Gamma(b+1)}+(-z)^{\frac{1}{2}+\alpha+\beta-b}\;
   \frac{\Gamma(c+1)\Gamma(a-b)}{\Gamma(c-b)\Gamma(a+1)}\Big]\Big\}.\label{4A.18}
\end{eqnarray}
To make the flux $\mathcal{F}$ vanish at the origin, we should have  poles of gamma functions
\begin{eqnarray}
   c-a=-n,\quad \textrm{or}\quad b+1=-n ,&\qquad\textrm{for}& \ell>0\label{4A.19}\\
    c-b=-n,\quad \textrm{or}\quad a+1=-n ,&\qquad\textrm{for}& \ell<0\label{4A.20}
\end{eqnarray}
Thus the quasinormal frequencies are
\begin{eqnarray}\label{4A.21}
\omega=-m-\frac{i}{l}(2n+\frac{3}{2}+\ell),&\quad \textrm{or}
\quad& \omega=m-\frac{i}{l}(2n+\frac{1}{2}+\ell),\qquad\textrm{for} \;\ell>0\label{4A.21}\\
\omega=-m-\frac{i}{l}(2n+\frac{1}{2}-\ell),&\quad \textrm{or}
\quad&
\omega=m-\frac{i}{l}(2n-\frac{1}{2}-\ell),\qquad\textrm{for}\;
\ell<0\label{4A.22}
\end{eqnarray}

From (\ref{4A.16})-(\ref{4A.18}), we know that when $\ell=0$ the
flux would not vanish at the origin, in contradictory to the QNM's
definition. So it is true for the fermion perturbation as well as
the scalar perturbation that there is no QNMs for the s-wave in
the three dimensional pure dS space.

\subsection{Four dimensional case}

We now turn our discussion to the four dimensional case. The
metric and the Dirac equation are showed as (\ref{2.1}) and
(\ref{4A.2}) but with $d=4$. In this spherical coordinate, a
general formalism was provided in \cite{20}(see also \cite{24}).
We will follow this setup. With the separation
\begin{eqnarray*}
    \Psi=\frac{e^{-i\omega t}}{\sqrt[4]{f}}\begin{pmatrix}
      \frac{iG^{\pm}(r)}{r}\varphi^{\pm}_{jm}(\theta,\phi) \\
      \frac{F^{\pm}(r)}{r}\varphi^{\mp}_{jm}(\theta,\phi) \\
    \end{pmatrix},
\end{eqnarray*}
where $\varphi^{\pm}_{jm}$ are the two component spinors with
$j=l\pm \frac{1}{2}$  \cite{23}, we arrive at the radial part
equation . If we use a new variable $r=l\sin\mu$, the equations
are like
   \begin{eqnarray}
  (\partial_\mu+\frac{\kappa_\pm}{\sin\mu})G^\pm &=& -(m l-\frac{\omega l}{\cos\mu})F^\pm ,\label{4B.1}\\
  (\partial_\mu-\frac{\kappa_\pm}{\sin\mu})F^\pm &=& -(m l+\frac{\omega
  l}{\cos\mu})G^\pm ,\label{4B.2}
\end{eqnarray}
where $\kappa_+$ and $\kappa_-$ are positive and negative
integers. These equations resemble equations
(\ref{4A.7})(\ref{4A.8}) in the form, if one replaces $i\omega
l,\ell,\Psi_1,\Psi_2$ there with $\kappa_\pm,i\omega
l,G^\pm,F^\pm$ here. So the general solutions are similar.
But things are a bit complicate here. We cannot find the purely
outgoing solutions as we did in the three dimensional case. Thus
we first consider the flux along the radial direction at the
origin $r=0$(or equivalently $z=0$ here)
\begin{eqnarray}\label{4B.3}
   &&
   \mathcal{F}=\sqrt{1-z}\sin\theta\;(|G^\pm|^2|\varphi_{jm}^\pm|^2-|F^\pm|^2|\varphi_{jm}^\mp|^2)
   \nonumber\\
   &&\qquad\simeq
   \sin\theta\;(|\psi_1^\pm|^2|\varphi_{jm}^\pm|^2-|\psi_2^\pm|^2|\varphi_{jm}^\mp|^2).
\end{eqnarray}
We can find the solutions that make the flux vanish at the origin.
For the case ($G^+,F^+,\kappa_+$), they are
\begin{eqnarray}
  &&  \psi_1^+=B_1^+ \;z^{\frac{1+\kappa_+}{2}}(1-z)^\beta\;_2F_1(a_1,b_1,c_1,z), \label{4B.4}\\
  &&
  \psi_2^+=B_2^+\;z^{\frac{\kappa_+}{2}}(1-z)^{\beta}\Big[\Big(\frac{1}{2}+\kappa_+-(\frac{1+\kappa_+}{2}-\frac{i\omega l}{2}+\beta)z\Big)\;_2F_1(a_1,b_1,c_1,z)\nonumber\\
  &&\qquad\qquad+\frac{a_1b_1}{c_1}z(1-z)\,\,_2F_1(a_1+1,b_1+1,c_1+1,z)\Big],\label{4B.5}
\end{eqnarray}
where the hypergeometric parameters $a,b$ and $c$ and the
relationship between the coefficients $B_1^+,B_2^+$ are
\begin{eqnarray*}
   && a_1=\frac{\kappa_++1-i\omega l}{2}+\beta
   ,\;\;b_1=\frac{2+\kappa_++i\omega
   l}{2}+\beta,\;\;
     c_1=\kappa_++\frac{3}{2},\\
     && B_1^+=B_2^+(\frac{\kappa_+}{2}+\frac{i\omega
     l}{2}\pm\beta).
\end{eqnarray*}

 For the case ($G^-,F^-,\kappa_-$), they are
\begin{eqnarray}
  &&  \psi_1^-=B_1^- \;z^{-\frac{\kappa_-}{2}}(1-z)^\beta\;_2F_1(a_2,b_2,c_2,z) \label{4B.6},\\
  &&
  \psi_2^-=B_2^-\;z^{\frac{1-\kappa_-}{2}}(1-z)^{\beta}\Big[(-\frac{1+\kappa_-}{2}+\frac{i\omega l}{2}-\beta)\;_2F_1(a_2,b_2,c_2,z)\nonumber\\
  &&\qquad\qquad+\frac{a_2b_2}{c_2}(1-z)\,\,_2F_1(a_2+1,b_2+1,c_2+1,z)\Big].\label{4B.7}
\end{eqnarray}
where the hypergeometric parameters  and coefficients are
\begin{eqnarray*}
   && a_2=-\frac{\kappa_-+i\omega l}{2}+\beta
   ,\;\;b_2=\frac{1-\kappa_-+i\omega
   l}{2}+\beta,\;\;
     c_2=-\kappa_-+\frac{1}{2},\\
     && B_1^-=B_2^-(\frac{\kappa_-}{2}+\frac{i\omega
     l}{2}\pm\beta).
\end{eqnarray*}
Note that $\beta$ is the solution of equation
$\beta^2+\frac{1}{4}(m l+\frac{i}{2})^2=0$.

To obtain the QNMs, we require that the waves are purely outgoing
at the horizon. In other words, the wave function is of the form
$(-\frac{1}{z})^{-\frac{i\omega l}{2}}$ around the horizon
$(z\rightarrow -\infty)$. To eliminate the ingoing part in the
wave function
 $\psi_{1,2}$, again, we appeal to the property of
hypergeometric function illustrated in (\ref{4A.17}). The incoming
and outgoing parts are then detached which enable us to remove the
incoming one thoroughly by setting poles, which are
\begin{eqnarray}
  c_1-b_1=-n,\;\;\textrm{or}\;\; a_1+1=-n, && \qquad \textrm{for}\;\;
  (G^+,F^+,\kappa_+); \label{4B.8}\\
c_2-b_2=-n,\;\;\textrm{or}\;\; a_2+1=-n, && \qquad
\textrm{for}\;\;
  (G^-,F^-,\kappa_-). \label{4B.9}
\end{eqnarray}
The corresponding frequencies of  QNMs are obtained as:

For ($G^+,F^+,\kappa_+$)
\begin{eqnarray}
   \omega=-m-\frac{i}{l}(2n+\kappa_++\frac{3}{2}),&&\quad\textrm{or} \nonumber \\
  \omega=m-\frac{i}{l}(2n+\kappa_++\frac{1}{2}).&& \label{4B.10}
\end{eqnarray}

For ($G^-,F^-,\kappa_-$)
\begin{eqnarray}
   \omega=-m-\frac{i}{l}(2n-\kappa_-+\frac{1}{2}),&&\quad\textrm{or}\nonumber\\
  \omega=m-\frac{i}{l}(2n-\kappa_--\frac{1}{2}).&& \label{4B.11}
\end{eqnarray}

These results are very similar to that of three dimensional case,
showing  consistent behavior of the fermion perturbation
in pure dS space.

\section{Conclusions and discussions}

We have studied the QNMs of the scalar and fermion perturbations in
the pure dS space. For the scalar perturbations, we have confirmed
the argument(see \cite{14}) that no QNM exists for the massless
scalar perturbations in four dimensional space or, more generally
, in arbitrary dimensional case.
To allow the existence of QNMs of scalar perturbation,
we have found that there is a
constraint on the mass of scalar field, that is
$m>\frac{d-1}{2l}$. Moreover, we have found that in the three
dimensional pure dS space QNM does not exist for the s-wave  scalar
perturbation. This is a special result in three dimensional case and does not exist in other dimensions.

We have also extended the discussion of scalar perturbation to the
topological dS spaces. In two special cases, the flat hypersurface
and the hyperbolic hypersurface, perturbations experience
amplification behaviors. Here again we see that topology influence
the behavior of the perturbation as that we observed in AdS
situation\cite{8}. In AdS cases, the wave amplification was only
found in the hyperbolic spaces \cite{8}, however in dS spaces the
growing modes are obtained in all topological spaces.  This serves
as an additional difference between the dS and AdS spacetimes.

 We have also investigated the fermion
perturbations in the three and four dimensional pure dS spaces.
The well-defined QNM frequencies were obtained. Similar to the
scalar perturbation, we again found that for three dimensional
case no QNMs exists for the s-wave perturbation.

Examining the QNM frequencies of both scalar and Fermion
perturbations, we found that though the real part of scalar
perturbation is proportional to the surface gravity of the
cosmological horizon, this dependence disappears in Fermion
perturbation. In additional to the difficulty to fully understand
the quantum connection between QNM and loop gravity
\cite{21}\cite{12}, it is also too early to see the relation
between the QNMs and thermodynamics of horizons.

ACKNOWLEDGEMENT: This work was partially supported by  NNSF of
China, Ministry of Education of China and Shanghai Science and
Technology Commission. R. K. Su's work was also supported by National Basic Research Programme of China. B. Wang would like to thank R. G. Cai for
helpful discussions.


\end{document}